\documentclass[aps,prd,onecolumn,groupedaddress,showpacs,nofootinbib,amssymb]{revtex4}
\usepackage[dvips]{graphicx}
\usepackage{amssymb}
\usepackage{amsmath}
\usepackage{graphicx}
\usepackage{amsfonts}
\usepackage{bm}

\begin{document}

\title{Inflation and Bounce from Classical and Loop Quantum Cosmology Imperfect Fluids}
\author{
V.K. Oikonomou,$^{1,2}$\,\thanks{v.k.oikonomou1979@gmail.com}}
\affiliation{
$^{1)}$ Laboratory for Theoretical Cosmology, Tomsk State University of Control Systems\\
and Radioelectronics (TUSUR), 634050 Tomsk, Russia \\
$^{2)}$ Tomsk State Pedagogical University, 634061 Tomsk  }

\begin{abstract}
We investigate how various inflationary and bouncing cosmologies can be realized by imperfect fluids with a generalized equation of state, in the context of both classical and loop quantum cosmology. With regards to the inflationary cosmologies, we study the intermediate inflation scenario, the $R^2$ inflation scenario and two constant-roll inflation scenarios and with regards to the bouncing cosmologies we study the matter bounce scenario, the singular bounce and the super bounce scenario. Within the context of the classical cosmology, we calculate the spectral index of the power spectrum of primordial curvature perturbations, the scalar-to-tensor ratio and the running of the spectral index and we compare the resulting picture with the Planck data. As we demonstrate, partial compatibility with the observational data is achieved in the imperfect fluid description, however none of the above scenarios is in full agreement with data. This result shows that although it is possible to realize various cosmological scenarios using different theoretical frameworks, it is not guaranteed that all the theoretical descriptions are viable.

\end{abstract}


\maketitle



\def\pp{{\, \mid \hskip -1.5mm =}}
\def\cL{\mathcal{L}}
\def\be{\begin{equation}}
\def\ee{\end{equation}}
\def\bea{\begin{eqnarray}}
\def\eea{\end{eqnarray}}
\def\tr{\mathrm{tr}\, }
\def\nn{\nonumber \\}
\def\e{\mathrm{e}}

\section{Introduction}

Admittedly, one of the most fascinating observations the last ten
years was the direct detection of gravitational waves, but as some
of the most influential minds in gravitational waves admit, the
gravitational waves were expected to be directly observed since
their indirect observation back in the early 90's. What was not
expected at all in astrophysics and cosmology was the observation of
late-time acceleration in the late 90's \cite{riess}. This
observation had set the stage for new research towards the
consistent modelling of this late-time acceleration era of our
Universe. It is conceivable that this late-time acceleration era
cannot be solely expressed by a perfect matter fluid, and therefore
this allows the use of more general cosmological fluids to be
potential candidates for the consistent description of the late-time
era, but also for other evolution eras of our Universe, like for
example the early-time acceleration era. Also it is known that the
effective equation of state of our Universe is believed to change at
late-times, taking values very close to $-1$, with $w=-1$ being
known as the phantom divide line. Obviously, such cosmic evolution
needs quite general forms of dark fluids in order to consistently
describe it. Nowadays, several cosmological frameworks make use of
imperfect fluids
\cite{inhomogen1,nojodineos1,inhomogen2,brevik1,brevik2,fluid1,fluid2,fluid3},
which in some cases are also viscous, and this description is an
alternative to modified gravity cosmological frameworks
\cite{reviews1,reviews1a,reviews2,reviews3,reviews4}. Particularly,
it is known that imperfect fluids can even describe a phantom
evolution without the need for a phantom scalar to drive the
evolution \cite{inhomogen1,inhomogen2}. Also, it has been shown in
Refs. \cite{fluid1,fluid2,fluid3} that early-time acceleration can
also be described by imperfect fluids. Moreover in a recent work
\cite{oikonomoufluid} it was demonstrated how singular inflationary
cosmology can be realized by imperfect fluids.

The purpose of this work is to investigate how several inflationary
and bouncing scenarios can be realized by imperfect fluids. We shall
use two different theoretical frameworks, namely classical cosmology
and Loop Quantum Cosmology (LQC) \cite{LQC1,LQC2,LQC3,LQC4,LQC5}
(see \cite{LQC3,LQC4} where the derivation of the Hamiltonian in LQC
was firstly derived to yield the modified Friedman equation, and
also see \cite{LQC5} for a recent derivation of the effective
Hamiltonian in LQC, which was derived by demanding repulsive
gravity, as in Loop Quantum Gravity). In both cases we shall
investigate which imperfect fluid can realize various inflationary
and bouncing cosmology scenarios. The inflationary cosmology
\cite{inflation1,inflation2,inflation2a,inflation2b,inflation2c,inflation3,inflation4,inflation5,inflation6}
and bouncing cosmology
\cite{brande1,bounce1,bounce1a,bounce1b,bounce2,bounce3,bounce4,bounce5,bouncehunt,matterbounce1,matterbounce2,matterbounce3,matterbounce4,matterbounce5,matterbounce6,matterbounce7}
are two alternative scenarios for our Universe evolution. In the
case of inflation, the Universe starts from an initial singularity
and accelerates at early times, while in the case of the bouncing
cosmology, the Universe initially contracts until it reaches a
minimum radius, and then it expands again. With regards to
inflation, we shall be interested in four different inflationary
scenarios, namely the intermediate inflation
\cite{intermediate1,intermediate2,intermediate3,intermediate4}, the
Starobinsky inflation
\cite{starobinsky,oikonomouodi1,oikonomouodi2,oikonomouodi3}, and
two constant-roll inflation scenarios \cite{staronew}. With regards
to bouncing cosmologies, we shall be interested in realizing several
well studied bouncing cosmologies, and particularly the matter
bounce scenario
\cite{matterbounce1,matterbounce2,matterbounce3,matterbounce4,matterbounce5,matterbounce6,matterbounce7},
the superbounce scenario \cite{superbounce1,superbounce2} and the
singular bounce
\cite{singularbounce1,singularbounce2,singularbounce3}.

As we already mentioned we shall use two theoretical frameworks,
that of classical cosmology and that of LQC. After presenting the
reconstruction methods for realizing the various cosmologies with
imperfect fluids, we proceed to the realization of the cosmologies
by using the reconstruction methods. In the case of classical
cosmology, we will calculate the power spectrum of primordial
curvature perturbations, the scalar-to-tensor ratio and the running
of the spectral index for all the aforementioned cosmologies, and we
compare the results to the recent Planck data \cite{planck}. The
main outcome of our work is that, although the cosmological
scenarios we study in this paper are viable in other modified
gravity frameworks, these are not necessarily viable in all the
alternative modified gravity descriptions. As we will demonstrate,
in some cases the resulting imperfect fluid cosmologies are not
compatible at all with the observational data, and in some other
cases, there is partial compatibility.

We need to note that the perturbation aspects in LQC are not
transparent enough and assume that there are no non-trivial quantum
gravitational modifications arising due to presence of
inhomogeneities. As it was shown in \cite{ref1}, a consistent
Hamiltonian framework does not allow this assumption to be true. The
perturbations issues that may arise in the context of the present
work, are possibly more related to some early works in LQC
\cite{ref2}, so any calculation of the primordial power spectrum
should be addressed as we commented above.

The main results of this work are the following: firstly, in our
attempt to realize the intermediate inflation scenario, the
Starobinsky inflation scenario and two constant-roll inflation
scenarios, by using imperfect fluids in the context of classical
cosmology, only the Starobinsky model and one of the two
constant-roll scenarios resulted to a spectral index of primordial
curvature perturbations compatible with the Planck 2015
observational data. However in all cases the scalar-to-tensor ratio
was incompatible with observations. Secondly, we performed the same
analysis for bouncing cosmologies, and specifically for the matter
bounce scenario, the superbounce and the singular bounce scenario.
As we demonstrated, only the singular bounce yielded a spectral
index compatible with the latest Planck data, but in this case too,
the scalar-to-tensor ratio was not compatible with the data. In all
cases, we found what is the imperfect fluid description in the
context of LQC, however, no perturbation analysis was performed for
the reasons we discussed earlier.

This paper is organized as follows: In section II we describe in
some detail the reconstruction methods for imperfect fluid cosmology
realization, in the cases of classical and LQC cosmology. In section
III we present the various inflationary and bouncing cosmologies and
we study how it is possible to realize these by classical and LQC
imperfect fluids. Finally, the conclusions follow in the end of the
paper.

\section{Cosmology with Classical and LQC Imperfect Fluids: General Formalism}

In this section we shall briefly present the theoretical framework of imperfect fluid cosmology, both in the classical and the LQC cases. We assume that the geometric background is a Friedmann-Robertson-Walker metric of the form,
\begin{equation}\label{metricformfrwhjkh}
\mathrm{d}s^2=-\mathrm{d}t^2+a^2(t)\sum_i\mathrm{d}x_i^2\, ,
\end{equation}
where $a(t)$ denotes as usual the scale factor. The LQC case is an extension of the classical description, in which case the classical equations of motion are modified by using the LQC effective Hamiltonian, which in effect modifies the Friedmann equations. In this section we describe in some detail how the LQC imperfect fluid framework is constructed, and also we demonstrate that it can lead to the classical imperfect fluid theory, if the classical limit is taken.

\subsection{Classical Imperfect Fluids}

We start of our analysis with the classical viscous fluid case, and for a detailed analysis the reader is referred to \cite{inhomogen1,inhomogen2,brevik1,brevik2,fluid1,fluid2,fluid3}. It is conceivable that the imperfect fluid cosmology is a limiting case of the viscous fluid case, and we show this explicitly now. Consider a viscous fluid with effective energy density $\rho$ and effective pressure
$p$, which are related to each other as follows,
\begin{equation}\label{energypressurerelation}
p=F(\rho)-B(a(t),H,\dot{H},...)\, ,
\end{equation}
where the function $B(a(t),H,\dot{H},...)$ stands for the bulk viscosity, and as it can be seen it is a function of the scale factor, the Hubble rate and of the higher derivatives of the Hubble rate. The function $F(\rho)$ is the homogeneous part of the effective equation of state (EoS) given in Eq. (\ref{energypressurerelation}), and the inhomogeneity in the EoS is introduced by the bulk viscosity, which needs to be positive in order for the entropy change to have a positive sign during irreversible processes \cite{fluid3}. The FRW equations for the metric (\ref{metricformfrwhjkh}) are,
\be
\label{JGRG11}
\rho =\frac{3}{\kappa^2}H^2 \, ,\quad p = -\frac{1}{\kappa^2}\left(3H^2 +
2\dot H\right)\, ,
\ee
The energy momentum tensor corresponding to the viscous fluid with EoS (\ref{energypressurerelation}) is given below,
\begin{equation}\label{viscousenergymom}
T_{\mu \nu}=\rho\,u_{\mu}\,u_{\nu}+\Big{(}F(\rho)+B(H,\dot{H},...)\Big{)}\Big{[}g_{\mu \nu}+u_{\mu}\,u_{\nu}\Big{]}\, ,
\end{equation}
where $u_{\mu}$ is the comoving four velocity, and its specific form for the metric (\ref{metricformfrwhjkh}) is $u_{\mu}=(1,0,0,0)$. The conservation of the energy and momentum, results to the following continuity equation for the effective energy density $\rho$,
\begin{equation}\label{rhodensity}
\dot{\rho}+3H(\rho+F(\rho))=3\,H\,B(H,\dot{H},...)\, .
\end{equation}
For the purposes of this paper, we shall consider cosmological evolutions which can be generated by a homogeneous imperfect fluid, in which case the EoS will be of the form $p=F(\rho)$, and as we will demonstrate, the function $F(\rho)$ will be of the form $F(\rho)=-\rho-f(\rho)$, and only in some LQC cases, it will have a general form $F(\rho)$. Given a cosmological evolution in terms of its Hubble rate, we can find in a straightforward way the EoS of the imperfect homogeneous or viscous fluid that can generate such a cosmological evolution. From the FRW equations (\ref{JGRG11}), the effective energy density and effective pressure of the fluid can be written as functions of the cosmic time in the following way,
\be
\label{C1}
\rho =f_\rho(t)\, , \quad p = f_p (t)\, .
\ee
Hence, if the first equation can be solved with respect to the cosmic time $t$, we get $t=f_\rho^{-1} \left(\rho \right)$, so by substituting this in the second equation of Eq. (\ref{C1}), the effective pressure can be expressed as a function of the effective energy density is $p = f_p \left( f_\rho^{-1} \left(\rho \right) \right)$. In the case we just described we assumed that the resulting function $f_p\circ f_\rho^{-1}$ depends solely on the effective energy density $\rho$, but this happens only in the imperfect fluid case. In general, the resulting function might also contain the scale factor and the Hubble rate, and this can happen for example in the cases that the scale factor or the Hubble rate are non-invertible functions of the cosmic time. In this paper however we focus on non-viscous imperfect fluid generated cosmologies, so we will not have deal with these issues.

\subsection{Loop Quantum Cosmology Imperfect Fluids}

The Loop Quantum Cosmology imperfect fluid formalism is an extension
of the classical formalism in the case that the Friedmann equations
are modified by taking into account holonomy effects. In the context
of LQC, the effecting Hamiltonian which describes the Universe is
\cite{LQC1,LQC2,LQC3,LQC4,LQC5},
\begin{equation}\label{effhamilt}
\mathcal{H}_{LQC}=-3V\frac{\sin^2(\lambda \beta)}{\gamma^2\lambda^2}+V\rho\, ,
\end{equation}
with $\gamma$ and $\lambda$ being the Barbero-Immirzi parameter and an arbitrary parameter with dimensions of length respectively. In addition, $V$ is the total volume of the Universe, which is $V=a(t)^3$, where $a(t)$ is the scale factor, and $\rho$ is the effective energy density of the matter fluid present, which in our case is the imperfect fluid. Using the Hamiltonian constraint $\mathcal{H}_{LQC}=0$, we obtain the following equation,
\begin{equation}\label{hamiltonianconstr}
\frac{\sin^2(\lambda \beta)}{\gamma^2\lambda^2}=\frac{\rho}{3}\, ,
\end{equation}
and in conjunction with the following anticommutation identity
\begin{equation}\label{anticom}
\dot{V}=\{V,\mathcal{H}_{LQC}\}=-\frac{\gamma}{2}\frac{\partial \mathcal{H}_{LQC}}{\partial \beta},
\end{equation}
we obtain the Friedmann equation in the case of LQC
\cite{LQC1,LQC2,LQC3,LQC4,LQC5},
\begin{equation}\label{holcor1}
H^2=\frac{\kappa^2\rho}{3}\left (1-\frac{\rho}{\rho_c}\right )\, .
\end{equation}
The parameter $\rho_c$ encompasses all the quantum effects of the theory, and the classical limit of the Friedmann equation (\ref{holcor1}) can be obtained if $\rho_c$ tends to infinity. The effective energy density of the imperfect fluid satisfies the continuity equation,
\begin{equation}\label{cont}
\dot{\rho}(t)=-3H\Big{(}\rho(t)+p(t) \Big{)}\, ,
\end{equation}
with $p(t)$ denoting as usual the effective pressure. Differentiating the holonomy corrected Friedmann equation (\ref{holcor1}), and using the continuity equation (\ref{cont}), we obtain,
\begin{equation}\label{eqnm}
\dot{H}=-\frac{\kappa^2}{2}(\rho+p)(1-2\frac{\rho}{\rho_c})\, .
\end{equation}
The above equation can be solved with respect to the pressure $p(t)$, which reads,
\begin{equation}\label{pr}
p(t)=-\rho-\frac{2\dot{H}}{\kappa^2(1-\frac{2\rho}{\rho_c})}\, ,
\end{equation}
A crucial feature of the LQC formalism is that in the limit $\rho_c\to \infty$, the Friedmann equations (\ref{holcor1}) and (\ref{eqnm}) yield the classical Friedmann equations. Indeed, in the limit $\rho_c\to \infty$, the equations (\ref{holcor1}) and (\ref{eqnm}) become,
\begin{equation}\label{classicaleqns}
\dot{H}=-\frac{\kappa^2}{2}(\rho+p),\,\,\,H^2=\frac{\kappa^2\rho}{3}\, ,
\end{equation}
which are identical to the classical equations (\ref{JGRG11}). Having equations (\ref{holcor1}) and (\ref{pr}) at hand, and also given the Hubble rate of a specific cosmological evolution, we can easily find which imperfect fluid can generate such an evolution. An important feature is that in the limit $\rho_c\to \infty$ the resulting EoS of the LQC imperfect fluid must be identical to the corresponding classical EoS which generates the same cosmological evolution. As we will demonstrate in the following sections, this indeed happens in all the examples we shall present. In order to find the LQC EoS of the imperfect fluid, it is necessary to find the explicit functional dependence of the energy density as a function of time. By solving the Friedmann equation (\ref{holcor1}) with respect to $\rho$ we obtain,
\begin{equation}\label{sol1}
\rho=\frac{\kappa ^2 \rho_c\pm \sqrt{-12 H^2 \kappa ^2 \rho_c+\kappa ^4 \rho_c^2}}{2 \kappa ^2}\, ,
\end{equation}
so there are two solutions, so this point is crucial in order to choose the correct one. In order to do so, we can use the feature that the loop quantum corrected quantities should be identical to the classical ones in the limit $\rho_c \to \infty$. The classical expression of the EoS $w_{eff}=p/\rho$ as a function of the cosmic time is,
\begin{equation}\label{claseos}
w_{eff}=-1-\frac{2\dot{H}}{3H^2}\, ,
\end{equation}
so the loop quantum corrected expression should be identical to this. The loop quantum corrected EoS as a function of time is,
\begin{equation}\label{eosdef}
w_{eff}=-1-\frac{\dot{H}}{3H^2}\pm \frac{\rho_c \dot{H}}{3 H^2 \sqrt{\rho_c \left(\rho_c-12 H^2\right)}}\, ,
\end{equation}
and therefore, in order for the expressions (\ref{eosdef}) and (\ref{claseos}) to be identical in the limit $\rho_c\to \infty$, the energy density as a function of the cosmic time should be,
\begin{equation}\label{rhodefcorrect}
\rho=\frac{\kappa ^2 \rho_c- \sqrt{-12 H^2 \kappa ^2 \rho_c+\kappa ^4 \rho_c^2}}{2 \kappa ^2}\, .
\end{equation}
Noticing the functional form of the effective pressure $p$, we can
see that the only term not being expressed as a function of the
energy density $\rho$, is the derivative of the Hubble rate
$\dot{H}$. So by finding this derivative as a function of the cosmic
time $t$, we can have the resulting EoS as a function of $\rho$, and
this can easily be done by inverting Eq. (\ref{rhodefcorrect}), and
this finding $t=t (\rho)$. It is conceivable however that the case
for which the function $\rho (t)$ is invertible corresponds to the
imperfect fluid scenario, and if it is not invertible the fluid has
bulk viscosity since the function $\dot{H}$ cannot be expressed as
an explicit function of the energy density $\rho$. Practically, the
equations (\ref{rhodefcorrect}) and (\ref{pr}) will enable us to
find the EoS of the imperfect fluid that generates a given
cosmological evolution. In the next sections we shall investigate
which imperfect fluids can generate various inflationary and
bouncing cosmologies.

Before we close this section we need to discuss an important issue,
with regard to the EoS in LQC. Particularly, an important property
which is needed in the analysis of LQC imperfect fluids, is that the
EoS does not change when one goes to LQC. This is an issue that
requires special attention, as it was discussed in \cite{ref3}. As
it was discussed in \cite{ref3}, the assumption we made about the
EoS, is true only when inverse volume modifications are ignored. If
only holonomy modifications are present, only then there is no
change in the EoS.

\section{Inflationary and Bounce Cosmology Models from Classical and LQC Imperfect Fluids}

\subsection{Inflationary Models}

In this section we shall employ the methods we described in the previous two sections in order to realize some well-known inflationary cosmologies. The first is the nearly $R^2$ inflationary evolution which is a quasi-de Sitter cosmology \cite{starobinsky}, in which case the Hubble rate and the scale factor read,
\begin{equation}\label{hubstar}
H(t)\simeq H_i-\frac{M^2}{6}\left ( t-t_i\right ),\,\,\,a(t)=e^{H_i t-\frac{M^2}{12}(t-t_i)^2}\, .
\end{equation}
where $t_i$ is the time instance that inflation starts and in addition $H_i$ is the Hubble rate at $t=t_i$. It can be shown that in the context of $F(R)$ gravity, the cosmological evolution can be realized by the $F(R)$ gravity of the form $F(R)=R+\frac{1}{6M^2}R^2$, since the equations of motion are,
\begin{equation}\label{takestwo}
\ddot{H}-\frac{\dot{H}^2}{2H}+\frac{M^2}{2}H=-3H\dot{H}\, ,
\end{equation}
and due to the fact that during the inflationary phase, the terms $\ddot{H}$ and $\dot{H}$ are neglected. Another inflationary solution we shall present and study is the so-called intermediate inflation scenario \cite{intermediate1,intermediate2,intermediate3,intermediate4}, in which case the scale factor is,
\begin{equation}\label{intermediatescale}
a(t)=e^{A\,t^f}\, ,
\end{equation}
with $A$ being a constant parameter and $0<f<1$. Let us see how these two inflationary cosmologies can be realized by classical and loop quantum cosmology imperfect fluids. We start off with the classical picture, in which case we shall calculate the slow-roll indices and the resulting observational inflation indices.

\subsubsection{Classical Imperfect Fluid Description}

We start our analysis with the intermediate inflation cosmology (\ref{intermediatescale}), and we shall assume that the EoS of the imperfect fluid that realizes this cosmology is of the form,
\begin{equation}\label{eosgeneral}
p=-\rho-f(\rho)\, ,
\end{equation}
so our task is to find the analytic form of the function $f(\rho )$. The classical Friedmann equations (\ref{JGRG11}) for the intermediate inflation yield,
\begin{equation}\label{rhoclas}
\rho=\frac{3 A^2 f^2 t^{-2+2 f}}{\kappa ^2}\, ,
\end{equation}
so by solving with respect to $t$ and substituting in the expression for the pressure, the pressure as a function of $\rho$ is,
\begin{equation}\label{ysiximousiki}
p=-\rho-B\,\rho ^{\frac{-2+f}{2 (-1+f)}}\, ,
\end{equation}
and therefore, the function $f(\rho )$ is equal to $f(\rho)=B\,\rho ^{\frac{-2+f}{2 (-1+f)}}$, where we set $B$ equal to,
\begin{equation}\label{beta}
B=2\ 3^{-\frac{-2+f}{2 (-1+f)}} A^{\frac{1}{-1+f}} (-1+f) f^{\frac{1}{-1+f}} \kappa ^{-\frac{f}{-1+f}}\, .
\end{equation}
By repeating the same procedure for the quasi-de Sitter cosmology of Eq. (\ref{hubstar}), the resulting EoS of the imperfect fluid that realizes this cosmology has a very simple form, which is,
\begin{equation}\label{eosnearlydesitter}
p=-\rho+\frac{M^2}{3 \kappa ^2}\, ,
\end{equation}
so in this case the function $f(\rho)$ is constant, that is, $f(\rho)=-\frac{M^2}{3 \kappa ^2}$. For the classical imperfect fluid cosmology we can calculate the inflationary observational indices, and particularly the
spectral index of primordial curvature perturbations $n_s$, the
scalar-to-tensor ration $r$ and the running of the spectral index $a_s$. We shall use the formalism of Ref. \cite{nojodineos1}, and as it was shown, the indices in terms of the function $f(\rho (N))$ and $\rho$, are equal to,
\begin{align}
n_\mathrm{s} - 1 =& - 9 \rho(N) f(\rho (N)) \left( \frac{\tilde
f'(\rho (N))-2}{2\rho (N)
- f(\rho (N))}\right)^2
+\frac{6\rho (N)}{2\rho (N) - f(\rho (N))}
\left\{ \frac{ f(\rho (N))}{\rho (N)} \right. \nn
& \left. + \frac{1}{2} \left(f'(\rho (N))\right)^2
+ f'(\rho (N)) -\frac{5}{2} \frac{f(\rho (N)) f'(\rho
(N))}{\rho (N)}
+ \left(\frac{f(\rho)}{\rho(N)}\right)^2+\frac{1}{3} \frac{\rho
'(N)}{f(\rho (N))}
\right. \nonumber \\
& \left.
\times \left[\left( f'(\rho (N)) \right)^2 + f(\rho (N))
f''(\rho (N))
-2 \frac{ f(\rho (N)) f'(\rho (N))}{\rho (N)}
+ \left( \frac{ f(\rho (N))}{\rho (N)} \right)^2 \right] \right\} \,,
\label{eq:2.32} \\
r =& 24\rho (N) f(\rho (N))
\left( \frac{ f'(\rho (N))-2}
{2\rho (N) - f(\rho (N))}\right)^2 \,,
\label{eq:2.33} \\
\alpha_\mathrm{s} =& \rho (N) f(\rho (N)) \left( \frac{\tilde
f'(\rho (N))-2}
{2\rho (N) - f(\rho (N))}\right)^2 \left[ \frac{72\rho (N)}{2\rho (N)
- f(\rho (N))} J_1 \right. \nn
& \left. -54 \rho (N) f(\rho (N)) \left( \frac{f'(\rho (N))-2}
{2\rho (N) - f(\rho (N))}\right)^2 -\frac{1}{f'(\rho
(N))-2}J_2 \right] \,,
\label{eq:2.34}
\end{align}
where $N$ is the e-folding number and the exact form of the functions $J_1$ and $J_2$ can be found in the Appendix. The latest Planck data \cite{planck} suggest that the values of the observational indices are constrained as follows,
\begin{equation}\label{planckconstr}
n_s=0.9644\pm 0.0049\, , \quad r<0.10\, , \quad a_s=-0.0057\pm 0.0071\, ,
\end{equation}
so we shall compare the observational data to values of the observational indices for the two imperfect fluids cases we found earlier. The calculation can be simplified in the case that the function $f(\rho (N))$ satisfies,
\begin{equation}\label{sda}
\frac{f(\rho )}{\rho }\ll 1\, ,
\end{equation}
a constraint that can be satisfied by both cases we presented earlier. In this case, the observational inflationary indices are simplified as follows \cite{nojodineos1},
\begin{equation}\label{insapprox}
n_s\simeq 1-6\frac{f(\rho )}{\rho (N)}\, , \quad r\simeq 24\frac{f(\rho
)}{\rho (N)}\, , \quad \alpha_s=-9\left(\frac{f(\rho )}{\rho
(N)}\right)^2\, .
\end{equation}
Having these at hand we can directly calculate the observational indices, which for the imperfect fluid EoS of Eq. (\ref{ysiximousiki}) are equal to,
\begin{equation}\label{interclass}
n_s\simeq \frac{4+f (-4+N)}{f N}\, , \quad r\simeq \frac{16 (-1+f)}{f N}\, , \quad \alpha_s=-\frac{4 (-1+f)^2}{f^2 N^2}\,
\end{equation}
where we used the fact that the function $\rho (N)$ in the case of the intermediate inflation is equal to,
\begin{equation}\label{intermediateinflrhon}
\rho (N)=\frac{3 A^{2-\frac{-2+2 f}{f}} f^2 N^{\frac{-2+2 f}{f}}}{\kappa ^2}\, .
\end{equation}
By using the values $(N,f)=(50,0.9)$ (recall that $0<f<1$), the observational indices become,
\begin{equation}\label{observindices}
n_s\simeq 1.00889,\,\,\,r\simeq -0.0355556,\,\, \,\alpha_s\simeq -0.0000197531,
\end{equation}
and as it can be seen the values do not satisfy any of the Planck constraints (\ref{planckconstr}). Equivalently, for the quasi-de Sitter case, the observational indices are,
\begin{equation}\label{quasidesitterobsrvind}
n_s=\simeq 1+\frac{2 M^2}{3 H_i^2-M^2 N+H_i M^2 t_i},\,\,\,r\simeq -\frac{8 M^2}{3 H_i^2-M^2 N+H_i M^2 t_i},\,\,\,\alpha_s\simeq -\frac{M^4}{\left(3 H_i^2-M^2 N+H_i M^2 t_i\right)^2}\, ,
\end{equation}
so by using the values $(N,H_i,t_i,M)=(60,10^{10},10^{-60},10^{10.5})$, the observational indices become,
\begin{equation}\label{observindicesnew1}
n_s\simeq 0.966499,\,\,\,r\simeq 0.134003,\,\, \,\alpha_s\simeq -0.000280577,
\end{equation}
and it can be seen that only the spectral index of primordial curvature perturbations agrees with the Planck data.

Another interesting inflationary scenario was studied in Ref. \cite{staronew}, and it involves some canonical scalar field models with a constant rate of slow-roll, to which we refer to as ``constant-roll inflationary models''. These models are phenomenologically interesting since these are similar to some well known inflationary scenarios in the literature. Particularly, one of these models is similar to the solution found in \cite{barrowconst}, while the other one is similar to hilltop inflation \cite{staronew}. In addition, these models have appealing observational features, since the observational indices in some cases are compatible with observational data. The main assumption of these scalar models is that there is a constant rate of roll, in which case the following relation holds true $\ddot{\varphi}/H\dot{\varphi}=-3-\alpha$, where the parameter $\alpha$ measures the deviation from the flat potential. We shall not describe in detail all the features of the models, since the theoretical framework is not related directly to our work, so we just use the resulting cosmological evolutions, which we shall realize by using classical and quantum imperfect fluids. Details on the scalar models we present here can be found in \cite{staronew}. The first inflationary model found in \cite{staronew}, has the following scale factor,
\begin{equation}\label{firstmodel}
a(t)=\sinh^{1/(3+\alpha)}\left((3+\alpha)M\,t\right)\, ,
\end{equation}
where the parameter $\alpha >-3$ and also $M$ is an integration constant which determines the amplitude of the power spectrum of curvature perturbations. The second inflationary solution has the following scale factor \cite{staronew},
\begin{equation}\label{secondmodel}
a(t)=\cosh^{-1/(3+\alpha)}\left((3+\alpha)M\,t\right)\, ,
\end{equation}
where in this case $\alpha<-3$. Using the imperfect fluid reconstruction method we described earlier, the resulting EoS which realizes the cosmology (\ref{firstmodel}) is equal to,
\begin{equation}\label{staroneweos1}
p=-\rho -\frac{6 M^2}{\kappa ^2}-\frac{2 M^2 \alpha }{\kappa ^2}+2\rho +\frac{2 \alpha  \rho }{3}\, .
\end{equation}
We can calculate the observational indices for this case, and their analytic form is,
\begin{align}\label{observstaro1}
& n_s=\simeq 13+4 \alpha -\frac{12 M^3 (3+\alpha )^2}{\kappa ^2 \arcsin\left(e^{N (3+\alpha )}\right)},\,\,\,r\simeq \frac{16 (3+\alpha ) \left(3 M^3 (3+\alpha )-\kappa ^2 \arcsin\left(e^{N (3+\alpha )}\right)\right)}{\kappa ^2 \arcsin\left(e^{N (3+\alpha )}\right)},
\\ \notag &
\alpha_s\simeq -\frac{4 (3+\alpha )^2 \left(-3 M^3 (3+\alpha )+\kappa ^2 \arcsin\left(e^{N (3+\alpha )}\right)\right)^2}{\kappa ^4 \arcsin\left(e^{N (3+\alpha )}\right)^2}\, ,
\end{align}
so by using $\alpha=-2.99$, $M\sim \kappa$ and $N=50$, the values of the observational indices are,
\begin{equation}\label{staronewresults1}
n_s\simeq 0.966913,\,\,\,r\simeq 0.132348,\,\, \,\alpha_s\simeq -0.000273689.
\end{equation}
At is can be seen, only the spectral index of primordial curvature perturbations is compatible with the Planck data, and this picture occurs by choosing other values of the parameters. Notice that as $\alpha$ approaches zero, the spectrum loses its red tilt and approaches the scale invariant value $n_s=1$. With regards to the second model (\ref{secondmodel}), the EoS which realizes this cosmological evolution is,
\begin{equation}\label{staroneweos1111}
p=- \rho +\frac{6 M^2}{\kappa ^2}+\frac{2 M^2 \alpha }{\kappa ^2}-\left(2 +\frac{2 \alpha  }{3}\right)\rho\, .
\end{equation}
The analytic expressions of the resulting observational indices for this case are,
\begin{align}\label{observstaro1111}
& n_s=\simeq -11-4 \alpha +\frac{12 M^3 (3+\alpha )^2}{\kappa ^2 \text{arcsech}\left(e^{N (3+\alpha )}\right)},\,\,\,r\simeq -\frac{16 (3+\alpha ) \left(3 M^3 (3+\alpha )-\kappa ^2 \text{arccosh}\left(e^{-N (3+\alpha )}\right])\right)}{\kappa ^2 \text{arccosh}\left(e^{-N (3+\alpha )}\right)},
\\ \notag &
\alpha_s\simeq -\frac{4 (3+\alpha )^2 \left(-3 M^3 (3+\alpha )+\kappa ^2 \text{arccosh}\left(e^{-N (3+\alpha )}\right)\right)^2}{\kappa ^4 \text{arccosh}\left(e^{-N (3+\alpha )}\right)^2}\, ,
\end{align}
so by using $\alpha=-3.00000001$, $M/ \kappa\simeq 10^{-2}$ and $N=60$, the values of the observational indices are,
\begin{equation}\label{staronewresults1111}
n_s\simeq 1,\,\,\,r\simeq -1.60014\times 10^{-14},\,\, \,\alpha_s\simeq -4.00071 \times 10^{-24},
\end{equation}
hence the resulting spectrum is exactly scale invariant.

In conclusion, in this section we have shown that many interesting inflationary scenarios coming from various theoretical contexts, can be realized by classical cosmological imperfect fluids, however even though these scenarios were originally viable in the context of their theoretical frameworks, their viability is not guaranteed in the imperfect fluid description.
\begin{table*}
\small
\caption{\label{classical} The Classical Imperfect Fluid Equations of State for Various Inflationary Cosmologies}
\begin{tabular}{@{}crrrrrrrrrrr@{}}
\tableline
\tableline
\tableline
Inflationary Model& Imperfect Fluid Equation of State
\\\tableline
$a(t)=e^{A\,t^f}$ & $p=-\rho-2\ 3^{-\frac{-2+f}{2 (-1+f)}} A^{\frac{1}{-1+f}} (-1+f) f^{\frac{1}{-1+f}} \kappa ^{-\frac{f}{-1+f}}\,\rho ^{\frac{-2+f}{2 (-1+f)}}$
\\\tableline
$a(t)=e^{H_i t-\frac{M^2}{12}(t-t_i)^2}$ & $p=-\rho+\frac{M^2}{3 \kappa ^2}$
\\\tableline
$a(t)=\sinh^{1/(3+\alpha)}\left((3+\alpha)M\,t\right)$ & $p=-\rho -\frac{6 M^2}{\kappa ^2}-\frac{2 M^2 \alpha }{\kappa ^2}+2\rho +\frac{2 \alpha  \rho }{3}$
\\\tableline
$a(t)=\cosh^{-1/(3+\alpha)}\left((3+\alpha)M\,t\right)$ & $p=- \rho +\frac{6 M^2}{\kappa ^2}+\frac{2 M^2 \alpha }{\kappa ^2}-\left(2 +\frac{2 \alpha  }{3}\right)\rho$
\\\tableline
\tableline
 \end{tabular}
\end{table*}
Hence, this shows that even though there can be many theoretical realizations of various cosmologies, the observational validity of the cosmological evolution cannot be guaranteed for all the theoretical modified gravity descriptions. In Table \ref{classical} we gathered all the results of this section, and we now proceed to the LQC imperfect fluid description of these inflationary cosmologies.

\subsubsection{LQC Imperfect Fluid Description}

In the case of LQC imperfect fluids, the procedure to obtain the EoS of the imperfect fluids that realize the inflationary cosmology we studied in the previous section is straightforward. So we study one of the cases in detail and the rest of the results can be found in Table \ref{lqcinflation}. Consider for example the intermediate inflation case (\ref{intermediatescale}), and by applying the LQC reconstruction method we presented in a previous section, we easily find that the function $\rho (t)$ given in Eq. (\ref{rhodefcorrect}) is equal to,
\begin{equation}\label{lqcrhointermediate}
\rho (t)=\frac{\rho_c}{2}-\frac{\sqrt{-12 A^2 f^2 t^{-2+2 f} \kappa ^2 \rho_c+\kappa ^4 \rho_c^2}}{2 \kappa ^2}\, ,
\end{equation}
so by inverting this, the function $t=t (\rho)$ is equal to,
\begin{equation}\label{trho}
t (\rho)=12^{-\frac{1}{-2+2 f}} \left(\frac{-4 \kappa ^4 \rho ^2+4 \kappa ^4 \rho  \rho_c}{A^2 f^2 \kappa ^2 \rho_c}\right)^{\frac{1}{-2+2 f}}\, ,
\end{equation}
and therefore by substituting this to the effective pressure of Eq. (\ref{pr}), the resulting imperfect fluid EoS is equal to,
\begin{equation}\label{lqcintermediateinfl}
p=-\rho -\frac{2\ 3^{-\frac{-2+f}{2 (-1+f)}} A (-1+f) f \rho_c \left(\frac{\kappa ^2 \rho  (-\rho +\rho_c)}{A^2 f^2 \rho_c}\right)^{\frac{-2+f}{2 (-1+f)}}}{\kappa ^2 (-2 \rho +\rho_c)}\, .
\end{equation}
An important feature that will validate our results is that the EoS of Eq. (\ref{lqcintermediateinfl}) should be identical to the classical EoS given in Eq. (\ref{ysiximousiki}), in the limit $\rho_c\to \infty$, and as it can be seen, the two expressions (\ref{lqcintermediateinfl}) and (\ref{ysiximousiki}) are identical in the limit $\rho_c\to \infty$. The rest of the LQC imperfect fluid equations of state that realize the inflationary cosmologies of Eqs. (\ref{hubstar}), (\ref{firstmodel}) and (\ref{secondmodel}), can be found in Table \ref{lqcinflation}.
\begin{table*}
\small
\caption{\label{lqcinflation} The LQC Imperfect Fluid Equations of State for Various Inflationary Cosmologies}
\begin{tabular}{@{}crrrrrrrrrrr@{}}
\tableline
\tableline
\tableline
Inflationary Model& Imperfect Fluid Equation of State
\\\tableline
$a(t)=e^{A\,t^f}$ & $p=-\rho -\frac{2\ 3^{-\frac{-2+f}{2 (-1+f)}} A (-1+f) f \rho_c \left(\frac{\kappa ^2 \rho  (-\rho +\rho_c)}{A^2 f^2 \rho_c}\right)^{\frac{-2+f}{2 (-1+f)}}}{\kappa ^2 (-2 \rho +\rho_c)}$
\\\tableline
$a(t)=e^{H_i t-\frac{M^2}{12}(t-t_i)^2}$ & $p=-\rho +\frac{M^2 \rho_c}{3 \sqrt{\kappa ^4 (-2 \rho +\rho_c)^2}}$
\\\tableline
$a(t)=\sinh^{1/(3+\alpha)}\left((3+\alpha)M\,t\right)$ & $p=-\rho -\frac{2 (3+\alpha ) \left(\kappa ^2 \rho  (\rho -\rho_c)+3 M^2 \rho_c\right)}{3 \kappa ^2 (-2 \rho +\rho_c)}$
\\\tableline
$a(t)=\cosh^{-1/(3+\alpha)}\left((3+\alpha)M\,t\right)$ & $p=-\rho +\frac{2 (3+\alpha ) \left(\kappa ^2 \rho  (\rho -\rho_c)+3 M^2 \rho_c\right)}{3 \kappa ^2 (-2 \rho +\rho_c)}$
\\\tableline
\tableline
 \end{tabular}
\end{table*}
By looking the resulting expressions for the equations of state appearing in Table \ref{lqcinflation}, these are identical to the corresponding ones in Table \ref{classical}, when the limit $\rho_c \to \infty$ is taken.

\subsection{Bouncing Models from Classical and LQC Imperfect Fluids}

The big bounce cosmology offers an appealing and promising alternative to the inflationary paradigm, since in the big bounce case the initial singularity is avoided. This is because the Universe never reaches the size zero, so effectively all crushing types of singularities are avoided. In this section we shall investigate how various bouncing models can be realized from classical and LQC imperfect fluids. In the case of classical imperfect fluids, we shall also check the viability of the resulting cosmology by calculating the observational indices. The models we shall investigate are: the well known matter bounce scenario \cite{matterbounce1,matterbounce2,matterbounce3,matterbounce4,matterbounce5,matterbounce6,matterbounce7}, the superbounce scenario \cite{superbounce1,superbounce2}, and finally the singular bounce \cite{singularbounce1,singularbounce2,singularbounce3}. The matter bounce scenario and the singular bounce scenario have quite appealing observational properties in the context of modified gravity and LQC, and one of the aims of this section is to directly check whether the observational viability of these models still occurs even in the context of imperfect fluid cosmology.

We start off the analysis with the matter bounce scenario, in which case the scale factor is,
\begin{equation}\label{matterbounceanalyticsals1}
a(t)=\left(\frac{3}{4} \kappa ^2\mu  t^2+1\right)^{\frac{1}{3 }}\, ,
\end{equation}
where $\mu$ is a mass scale. Employing the imperfect fluid reconstruction techniques we described in the previous sections, the resulting classical imperfect fluid EoS that realizes the matter bounce cosmology is,
\begin{equation}\label{matterbounceanalyticsals2}
p=\frac{1}{2} \left(\mu +\frac{\sqrt{\kappa ^4 \mu ^3 (\mu -4 \rho )}}{\kappa ^2 \mu }-4 \rho \right)\, .
\end{equation}
Accordingly, a direct calculation of the observational indices yields,
\begin{align}\label{matterbounceanalyticsals3}
& n_s=\simeq 1+\frac{9}{2} N^2 \kappa ^2 \mu,\,\,\,r\simeq -18 N^2 \kappa ^2 \mu,\,\,\,
\alpha_s\simeq -\frac{\left(16-4 \kappa +3 N^2 \kappa ^3 \mu +9 N^4 \kappa ^4 \mu ^2\right)^2}{64 N^4 \kappa ^4 \mu ^2}\, .
\end{align}
By looking the functional form of the observational indices, we can easily conclude that the resulting cosmology is not viable, since there is no way that the power spectrum can be scale invariant or even nearly scale invariant.

The second bouncing cosmology we shall investigate is the superbounce cosmology \cite{superbounce1,superbounce2}, in which case the scale factor is,
\begin{equation}\label{singularbouncenasjshdeg1}
a(t)=(t-t_s)^{2\left/c^2\right.}\, ,
\end{equation}
where $c$ is a parameter which satisfies $c>\sqrt{6}$ and $t_s$ the bouncing time instance. Using the imperfect fluid reconstruction method we described earlier, the resulting EoS which realizes the cosmology (\ref{singularbouncenasjshdeg1}) is very simple and has the following functional form,
\begin{equation}\label{singularbouncenasjshdeg2}
p=-\rho +\frac{c^2 \rho }{3}\, .
\end{equation}
We can calculate the observational indices for this case, and their analytic form is,
\begin{align}\label{singularbouncenasjshdeg3}
& n_s=\simeq 1+2 c^2,\,\,\,r\simeq -8 c^2,\,\,\,\alpha_s\simeq -c^4\, ,
\end{align}
Hence, in this case too, the observational features of the superbounce cosmology in the context of imperfect fluids are not appealing at all, since no scale invariant power spectrum occurs.
\begin{table*}
\small
\caption{\label{classicalbounce1} The Classical Imperfect Fluid Equations of State for Various Bouncing Cosmologies}
\begin{tabular}{@{}crrrrrrrrrrr@{}}
\tableline
\tableline
\tableline
Bouncing Model& Classical Imperfect Fluid Equation of State
\\\tableline
$a(t)=\left(\frac{3}{4} \kappa ^2\mu  t^2+1\right)^{\frac{1}{3 }}$ & $p=\frac{1}{2} \left(\mu +\frac{\sqrt{\kappa ^4 \mu ^3 (\mu -4 \rho )}}{\kappa ^2 \mu }-4 \rho \right)$
\\\tableline
$a(t)=(t-t_s)^{2/c^2}$ & $p=-\rho +\frac{c^2 \rho }{3}$
\\\tableline
$a(t)=e^{f_0(t-t_s)^{\alpha}}$ & $p=-\rho -2\ 3^{-\frac{-2+\alpha }{2 (-1+\alpha )}} f_0^{\frac{1}{-1+\alpha }} (-1+\alpha ) \alpha ^{\frac{1}{-1+\alpha }} \kappa ^{-\frac{\alpha }{-1+\alpha }} \rho ^{\frac{-2+\alpha }{2 (-1+\alpha )}}$
\\\tableline
\tableline
 \end{tabular}
\end{table*}

The most interesting case for bouncing cosmologies in the context of imperfect fluids is the singular bounce evolution, in which case the scale factor is,
\begin{equation}\label{superbouncgdfsfawegddsw1}
a(t)=e^{(t-t_s)^{\alpha}}\, ,
\end{equation}
where $\alpha>1$ and $t=t_s$ is the bouncing point which coincides with the singular point. The singular bounce is called singular because at the bouncing point a Type IV singularity occurs \cite{Nojiri:2005sx}. Before proceeding in the imperfect fluid realization of this cosmology, it is worth noticing that the singular bounce and the intermediate inflation scenario look quite similar, at least their function forms. However, the difference is that the singular bounce starts with a contraction, and the parameter $\alpha$ has to be $\alpha=2n/(2m+1)$, in order to avoid complex values in the scale factor and Hubble rate. In the intermediate inflation case, the evolution starts at $t=0$ and the initial singularity is a Type II singularity, since $0<f<1$. This issue deserves a more detailed analysis, so we defer this to a future work. Using the imperfect fluid reconstruction method, the resulting EoS which realizes the cosmology (\ref{superbouncgdfsfawegddsw1}) is the following,
\begin{equation}\label{superbouncgdfsfawegddsw2}
p=-\rho -2\ 3^{-\frac{-2+\alpha }{2 (-1+\alpha )}} f_0^{\frac{1}{-1+\alpha }} (-1+\alpha ) \alpha ^{\frac{1}{-1+\alpha }} \kappa ^{-\frac{\alpha }{-1+\alpha }} \rho ^{\frac{-2+\alpha }{2 (-1+\alpha )}}\, .
\end{equation}
The analytic forms of the observational indices are the following,
\begin{align}\label{superbouncgdfsfawegddsw3}
& n_s=\simeq 1-\frac{4 (-1+\alpha )}{ N \alpha },\,\,\,r\simeq \frac{16 (-1+\alpha )}{N \alpha } ,\,\,\,\alpha_s\simeq -\frac{4 (-1+\alpha)^2}{\alpha^2 N^2}\, ,
\end{align}
so by using $\alpha=2$ and $N=60$, the resulting values of the observational indices are,
\begin{equation}\label{superbouncgdfsfawegddsw4}
n_s\simeq 0.966667,\,\,\,r\simeq 0.133333,\,\, \,\alpha_s\simeq -0.000277778.
\end{equation}
So the resulting power spectrum is a red tilted scale invariant spectrum, however the scalar-to-tensor ratio is excluded from the Planck data (\ref{planckconstr}). Nevertheless this scenario is the only bounce cosmology which yields a nearly scale invariant power spectrum. In Table \ref{classicalbounce1} we gathered the resulting classical imperfect fluid equations of state.

In a similar way, by using the LQC cosmology theoretical framework, the resulting equations of state can be easily found for all the bouncing cosmologies we studied earlier, and we presented the results in Table \ref{classicalbounce2}, since the procedure is the same as in the inflationary case.
\begin{table*}
\small
\caption{\label{classicalbounce2} The LQC Imperfect Fluid Equations of State for Various Bouncing Cosmologies}
\begin{tabular}{@{}crrrrrrrrrrr@{}}
\tableline
\tableline
\tableline
Bouncing Model& LQC Imperfect Fluid Equation of State
\\\tableline
$a(t)=\left(\frac{3}{4} \kappa ^2\mu  t^2+1\right)^{\frac{1}{3 }}$ & $p=\frac{\kappa ^2 \mu  (\mu -4 \rho ) \rho_c+\sqrt{\kappa ^4 \mu ^3 (\mu -4 \rho ) \rho_c^2}}{2 \kappa ^2 \mu  (-2 \rho +\rho_c)}$
\\\tableline
$a(t)=(t-t_s)^{2/c^2}$ & $p=-\rho -\frac{c^2 \kappa ^2 \rho  (\rho -\rho_c)}{3 \sqrt{\kappa ^4 (-2 \rho +\rho_c)^2}}$
\\\tableline
$a(t)=e^{f_0(t-t_s)^{\alpha}}$ & $p=-\rho -\frac{2\ 3^{-\frac{-2+\alpha }{2 (-1+\alpha )}} f_0^{\frac{1}{-1+\alpha }} (-1+\alpha ) \alpha ^{\frac{1}{-1+\alpha }} \kappa ^{\frac{-2+\alpha }{-1+\alpha }} \rho ^{\frac{-2+\alpha }{2 (-1+\alpha )}} \rho_c^{\frac{\alpha }{2 (-1+\alpha )}} (-\rho +\rho_c)^{\frac{-2+\alpha }{2 (-1+\alpha )}}}{\sqrt{\kappa ^4 \left(4 \rho  (\rho -\rho_c)+\rho_c^2\right)}}$
\\\tableline
\tableline
 \end{tabular}
\end{table*}
As it can be seen in Table \ref{classicalbounce2}, the LQC imperfect fluid equations of state become identical to the corresponding classical equations of state, in the limit $\rho_c\to \infty$.

\section{Conclusions}

Motivated by the fact that the Universe has an equation of state
which is around the phantom divide line value $w=-1$, in this paper
we investigated how some well know inflationary and bouncing
cosmologies can be realized by imperfect fluids. Particularly, with
regards to the inflationary scenarios, we examined the intermediate
inflation scenario, the Starobinsky inflation scenario and two
constant-roll inflationary scenarios. With regards to the bouncing
cosmologies, we examined the matter bounce scenario, the superbounce
and the singular bounce scenario. We found the imperfect fluid
description for each of the aforementioned cosmologies, and non of
these is described by a viscous fluid. We used two theoretical
frameworks for the imperfect fluids, the classical cosmology
framework and the LQC framework. In the case of the classical
cosmology, and with regards to the inflationary scenarios, the
Starobinsky inflationary scenario and one of the constant-roll
scenarios resulted to a spectral index of primordial curvature
perturbations compatible with the observational data, however the
rest of the indices were not compatible with the observations. With
respect to the bouncing cosmologies, only the singular bounce
yielded a spectral index compatible with the latest Planck data, but
in this case too, the rest of the indices were not compatible with
the data.

Hence the present article leads to the conclusion that not all
theoretical frameworks lead to successful descriptions of the
Universe. The main feature that renders a theoretical description
viable is the compatibility with the observational data, and as we
showed the imperfect fluid classical cosmology description provides
partial compatibility with data, and therefore it is somehow
incomplete. Note that the Starobinsky model in the context of $F(R)$
modified gravity is a viable model, and therefore the imperfect
fluid case does not make it viable. However, we need to note that we
have not extended the calculation of the slow-roll indices for an
imperfect fluid in the context of LQC. Possibly in this case the
holonomy corrected theory has something new to offer to the
imperfect fluid theoretical description. This task is a non-trivial
exercise, which we intend to do in a future work.

\section*{Acknowledgments}

This work is supported by Min. of Education and Science of Russia (V.K.O).

\section*{Appendix: The Functions $J_1$ and $J_2$}

In this Appendix we provide the detailed form of the functions $J_1$ and
$J_2$, appearing in Eq.~(\ref{eq:2.34}). Their detailed form is \cite{oikonomoufluid},
\begin{align}
J_1 =& \frac{f(\rho (N))}{\rho (N)} + \frac{1}{2}
\left(f'(\rho (N))\right)^2 + f'(\rho (N)) 
-\frac{5}{2} \frac{f(\rho (N)) f'(\rho (N))}{\rho (N)}
+ \left(\frac{f(\rho (N))}{\rho (N)}\right)^2 \nonumber+\frac{1}{3}
\frac{\rho '(N)}{f(\rho (N))} \nn
&
\times \left[\left(f'(\rho (N))\right)^2 + f(\rho (N)) \tilde
f''(\rho (N))
-2 \frac{f(\rho (N)) f'(\rho (N))}{\rho (N)} + \left(
\frac{f(\rho (N))}{\rho (N)}
\right)^2 \right]\,, \label{eq:2.35} \\
J_2 = &
\frac{45}{2} \frac{f(\rho (N))}{\rho (N)} \left(f'(\rho (N))
- \frac{1}{2} \frac{f(\rho (N))}{\rho (N)}\right) 
+ 18\left(\frac{f(\rho (N))}{\rho (N)}\right)^{-1} \left\{
\left(f'(\rho (N))-\frac{1}{2} \frac{ f(\rho (N))}{\rho
(N)}\right)^2 \right. \nn
& \left. + \left( f'(\rho (N))-\frac{1}{2} \frac{f(\rho
(N))}{\rho (N)}\right)^3 \right\} 
-9\left( f'(\rho (N)) - \frac{1}{2} \frac{f(\rho (N))}{\rho
(N)}\right)^2
-45 f'(\rho (N)) + 9\frac{f(\rho (N))}{\rho (N)}
\nonumber \\
&
+3 \left(4 f'(\rho (N)) -7\frac{ f(\rho (N))}{\rho (N)}
+2\right)
\left\{
-\frac{3}{2}\left(f'(\rho (N)) -\frac{1}{2}\frac{f(\rho
(N))}{\rho (N)}\right)
+ \left(\frac{f(\rho (N))}{\rho (N)}\right)^{-2}
\frac{\rho '(N)}{\rho (N)} \right. \nn
& \times \left[ \left( f'(\rho (N))\right)^2 + f(\rho (N))
f''(\rho (N)) 
\left.
- 2 \frac{ f(\rho (N)) f'(\rho (N))}{\rho (N)} + \left(
\frac{f(\rho (N))}{\rho (N)}
\right)^2 \right] \right\}
\nonumber \\
& +\left(\frac{ f(\rho (N))}{\rho (N)}\right)^{-2}
\left\{
-\frac{3}{2} \left(\frac{f(\rho (N))}{\rho (N)}\right)
\left(\frac{\rho '(N)}{\rho (N)}\right)
\left[
3\left( f'(\rho (N)) \right)^2
+2 f(\rho (N)) f''(\rho (N)) \right. \right. \nn
& \left. -\frac{11}{2}\frac{ f(\rho (N)) f'(\rho (N))}{\rho
(N)}
+\frac{5}{2} \left( \frac{f(\rho (N))}{\rho (N)} \right)^2
\right]
\nonumber \\
& \left.
+ \left(\frac{\rho ''(N)}{\rho (N)}\right)
\left[
\left( f'(\rho (N)) \right)^2 + f(\rho (N)) f''(\rho
(N))
-2\frac{ f(\rho ) f'(\rho )}{\rho (N)}
+ \left( \frac{ f(\rho )}{\rho } \right)^2
\right]
\right.
\nonumber \\
& \left.
+ \left(\frac{\rho '(N)}{\rho (N)}\right)^2
\left[ \left(3f'(\rho (N)) f''(\rho (N)) + f(\rho (N))
f'''(\rho (N)) \right) \rho (N)
-3\left( f'(\rho (N)) \right)^2 \right. \right. \nn
& \left. \left. - 3 f(\rho (N)) f''(\rho (N))
+6\frac{f(\rho (N)) f'(\rho (N))}{\rho (N)}
-3\left( \frac{f(\rho (N))}{\rho (N)} \right)^2
\right]
\right.
\,.
\label{eq:2.36}
\end{align}

\end{document}